\documentclass[sigconf]{acmart}
\usepackage{multirow}
\usepackage{enumitem}
\usepackage[most]{tcolorbox}

\setcopyright{none}                
\begin{document}

\title{\textsc{TraceSynth}: Generating Production-Quality Kernel Traces with Constraint-Guided Diffusion Models}
\titlenote{Author's accepted version. The Version of Record appears in the \emph{34th ACM Joint
European Software Engineering Conference and Symposium on the Foundations of Software Engineering
(FSE Companion '26)}, July 5--9, 2026, Montreal, QC, Canada.
\url{https://doi.org/10.1145/3803437.3805222}. Licensed under CC BY 4.0.}

\author{Yuvraj Sehgal}
\email{ys19rk@brocku.ca}
\affiliation{
  \institution{Department of Computer Science, Brock University}
  \city{St. Catharines}
  \state{Ontario}
  \country{Canada}
}

\author{Sneh Patel}
\email{sp18oo@brocku.ca}
\affiliation{
  \institution{Department of Computer Science, Brock University}
  \city{St. Catharines}
  \state{Ontario}
  \country{Canada}
}

\author{Mahsa Panahandeh}
\email{mpanahan@uottawa.ca}
\orcid{0000-0002-6369-8982}
\affiliation{
  \institution{School of Electrical Engineering and Computer Science, University of Ottawa}
  \city{Ottawa}
  \state{Ontario}
  \country{Canada}
}

\author{Naser Ezzati-Jivan}
\email{nezzatijivan@brocku.ca}
\orcid{0000-0003-1435-6297}
\affiliation{
  \institution{Department of Computer Science, Brock University}
  \city{St. Catharines}
  \state{Ontario}
  \country{Canada}
}

\author{Francois Tetreault}
\email{ftetreau@ciena.com}
\affiliation{
  \institution{Ciena Corporation}
  \city{Ottawa}
  \country{Canada}
}

\begin{abstract}
Machine learning models for system diagnostics rely on kernel execution traces to capture fine-grained system behavior, but collecting production traces in industrial systems is costly due to runtime overhead, storage demands, and privacy constraints.
We present \textsc{TraceSynth}, a diffusion-based framework for generating synthetic kernel traces that augment limited real data for downstream ML tasks.
\textsc{TraceSynth} models traces as multi-channel sequences (event types, timestamps, CPU affinity, thread identifiers, and process metadata) using a Transformer-based denoising diffusion process with constraint-guided repair to enforce system invariants.
Across six benchmarks, results show strong workload dependence.
For deterministic, compute-heavy workloads (\texttt{scimark2}), synthetic augmentation achieves 87.2\% F1-Macro at context length $L=4096$, only 2.6 percentage points below real-only baselines.
Context length is the dominant quality factor, with $L=4096$ yielding a +104\% relative improvement over $L=256$, while constraint-guided repair improves synthetic data quality by up to 4.3\%.
Ablation studies show that lightweight 2-channel models retain 97--99\% of the performance of full 6-channel models at roughly half the computational cost.
\textsc{TraceSynth} supports cost-effective augmentation of kernel execution traces in production observability pipelines and helps identify when synthetic data can substitute for limited real traces.
\end{abstract}



\maketitle

\section{Introduction}
\label{sec:intro}

Modern production systems increasingly rely on machine learning (ML) for operational intelligence, utilizing predictive models to forecast bottlenecks and automated analysis to reduce mean time to resolution (MTTR). To move beyond simple thresholding, these ML tools require high-fidelity data that captures the system's internal state rather than just its external symptoms. Kernel-level traces serve as this definitive source of truth. Unlike aggregate metrics, traces capture scheduler decisions, memory allocations, and I/O operations with microsecond-level precision, providing the rich structural data needed to train robust diagnostic models.

\noindent\textbf{The Data Scarcity Problem.} Collecting production kernel traces is expensive and constrained by several factors. First, while modern tracers like LTTng~\cite{Desnoyers2006The} provide high-throughput instrumentation, enabling them in production introduces non-trivial runtime overhead (typically 1.5--1.6$\times$~\cite{Darche2024Low-Overhead}), making continuous collection impractical for latency-sensitive services. Second, kernel traces often contain sensitive information—including file paths and network endpoints—that violate data retention policies. Third, real traces rarely capture the ``long-tail'' failure modes valuable for training.

This challenge is exemplified at Ciena, a global telecommunications provider. For Ciena, continuously instrumenting high-throughput network elements is operationally infeasible due to performance overhead. Furthermore, strict privacy policies regarding customer network topology prevent the aggregation of large-scale datasets for centralized training. Even when traces are captured during incidents, isolating the relevant execution segments remains a labor-intensive bottleneck.

As a result, practitioners face a fundamental tension: ML models require diverse and representative training data, yet production constraints limit what can be safely and cost-effectively collected. In practice, teams often resort to opportunistic trace collection during maintenance windows or from staging environments, yielding datasets that poorly reflect production diversity.

\noindent\textbf{Limitations of Existing Approaches.} Prior work on trace synthesis has largely focused on application-level logs~\cite{Gholamian2021A} or network traffic~\cite{bojchevski2018netgangeneratinggraphsrandom}, which lack the fine-grained system state captured in kernel traces. Statistical models such as Markov chains can generate locally valid event sequences but fail to capture long-range temporal dependencies~\cite{Hosoya1997A} and multi-attribute correlations inherent in kernel execution. Rule-based generators require substantial domain expertise and do not generalize easily across workloads~\cite{10764862}. Consequently, existing tools cannot generate synthetic traces that are both statistically diverse and structurally valid enough to replace real data.

\noindent\textbf{Our Approach.} We present \textsc{TraceSynth}, a framework for generating synthetic kernel execution traces using Denoising Diffusion Probabilistic Models (DDPMs)~\cite{ho2020denoisingdiffusionprobabilisticmodels}. Unlike recent work from our group~\cite{11029922} that focused on trace \textit{imputation} (reconstructing missing events), \textsc{TraceSynth} addresses the more challenging problem of \textit{unconditional generation}—creating complete, novel execution scenarios from scratch. To ensure generated traces respect system invariants—such as valid life-cycle transitions and correct CPU affinities—we introduce a constraint learning module that repairs output violations post-generation.

By enabling the synthesis of rare and complex execution patterns, \textsc{TraceSynth} allows engineers at Ciena and other industrial companies not only to train diagnostic models but also to stress-test trace-driven infrastructure against edge cases that are costly or impractical to reproduce in production.

\noindent\textbf{Contributions.} This paper makes the following contributions:
\begin{itemize}[leftmargin=*,nosep]
  \item \textbf{C1: Problem Characterization.} We define the data scarcity challenge for ML-based diagnostics at Ciena.
  \item \textbf{C2: \textsc{TraceSynth} Framework.} We propose a multi-channel diffusion model that synthesizes six trace attributes (event, time, CPU, thread, command, return) up to context length $L=4096$.
  \item \textbf{C3: Empirical Evaluation.} We demonstrate that synthetic augmentation effectively doubles training data size for deterministic workloads.
  \item \textbf{C4: Practical Guidance.} We provide actionable insights on cost-quality trade-offs, showing that simpler 2-channel models often suffice for industrial use cases.
\end{itemize}

\noindent\textbf{Paper Organization.}
Section~\ref{sec:related-works} discusses related work. Section~\ref{sec:approach} details \textsc{TraceSynth}. Sections~\ref{sec:evaluation} and~\ref{sec:results} present the setup, experiments and findings. Section~\ref{sec:discussion} discusses the implications and limitations. Finally, section~\ref{sec:conclusion} covers conclusion and future work.

\section{Related Work}
\label{sec:related-works}

Our work builds on three research areas: synthetic trace generation, constraint-guided generation, and diffusion models for structured sequences.

\subsection{Synthetic Trace Generation}

Kernel execution traces provide comprehensive visibility into system behavior, including scheduling, memory, and I/O operations~\cite{Noferesti2024Enhancing, Njoku2025Kernel-Level, Gebai2018Survey, Naas2021EZIOTracer, 9610601}. However, due to data scarcity, synthetic generation has become a critical research target.
Early approaches relied on pattern-matching and state-based abstraction. Ezzati-Jivan et al.~\cite{Ecole4861} proposed stateful event generation from kernel traces, converting platform-specific events into semantic abstractions. 
More recently, domain-specific methods have emerged: SAGA~\cite{Huang2024SAGASA} generates labeled audit logs for cyber-security using attack pattern templates. TraceLLM~\cite{kim2025largelanguagemodelsrealistic} applies large language models to microservice call graph generation through recursive layer-by-layer generation with instruction tuning.

\subsection{Constraint-Guided Generation}

Enforcing structural and semantic constraints remains a key challenge. Tracyn~\cite{huang2025projecttracyngenerativeartificial} reformulates PCIe trace generation as image generation and applies post-hoc constraint repair through dispersion-based calibration, achieving $1000\times$ improvement in task-specific metrics. This two-stage approach—unconstrained generation followed by targeted repair—demonstrates that post-hoc correction can be more effective than constrained generation. Related work has similarly shown that synthetic trace generation pipelines benefit from decoupling generation from validation and repair, particularly when realistic constraints are difficult to encode apriori~\cite{10764862}. Our constraint-guided repair mechanism follows this paradigm but learns constraints directly from real traces rather than relying on domain-specific rules.

\subsection{Diffusion Models for Structured Sequences}

Recent work has explored diffusion models for trace reconstruction. Janecek et al.~\cite{11029922} apply diffusion models (DiffWave, SSSDS4, CSDIS4) to reconstruct missing events in kernel traces caused by buffer overflows. Their SSSDS4 model achieves 81.62\% accuracy by leveraging surrounding context, significantly outperforming LSTM baselines. This establishes diffusion models as viable alternatives to traditional abstraction techniques.

Our work extends these foundations in three ways: (1) we generate complete multi-attribute traces rather than reconstructing missing segments, (2) we jointly model six correlated trace attributes in a unified latent space, and (3) we evaluate synthetic data quality through downstream task performance rather than reconstruction accuracy.

Toemmel et al.~\cite{toemmel2021catchganusingartificial} demonstrate that sequence-based GAN variants (e.g., SeqGAN, MaliGAN, CoT) fail for log generation due to their inability to maintain semantic properties. Despite achieving syntactic correctness, evaluated GANs violated chronology and event coherence. These findings highlight fundamental limitations of GAN-based sequence modeling~\cite{goodfellow2020generative},
motivating exploration of alternative generative paradigms—including
diffusion models—better suited to structured sequence modeling with hard constraints.

Unlike prior work focusing on specific domains (cyber-security logs, microservice traces, PCIe packets), we target general-purpose kernel execution traces with diverse workload characteristics. Our multi-channel diffusion model jointly generates correlated attributes while our constraint-guided repair ensures full structural validity.

\section{Approach}
\label{sec:approach}
\textsc{TraceSynth} addresses the execution trace data scarcity problem identified in Section~\ref{sec:intro}, instantiated in this work using kernel traces, by generating invariant-preserving synthetic traces that can be used in existing trace-driven analysis pipelines. The framework is designed to operate under realistic engineering constraints, including limited access to production traces and the need for predictable runtime and storage overhead. Across its stages, \textsc{TraceSynth} prioritizes semantic correctness and safe downstream reuse for training diagnostic and trace-driven models, rather than model-specific optimizations.

\begin{figure*}[t]
  \centering
  \includegraphics[width=\textwidth]{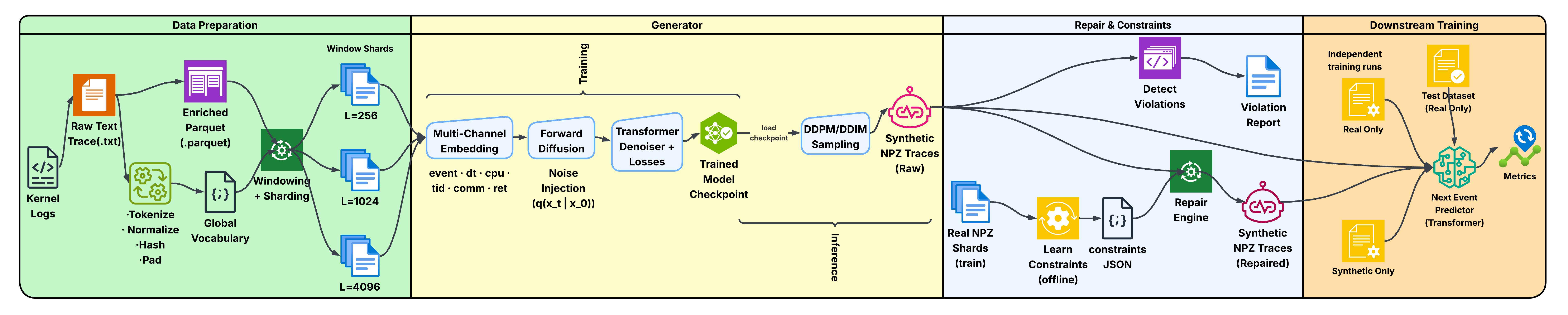}
\caption{\textsc{TraceSynth} end-to-end pipeline. Raw kernel traces are transformed into structured Parquet files and windowed NPZ shards at multiple context lengths. A multi-channel diffusion model is trained to capture recurring execution patterns across trace attributes. At inference time, the model generates synthetic traces, which are validated and repaired using learned system constraints before downstream evaluation.}
\label{fig:architecture}
\end{figure*}

Figure~\ref{fig:architecture} presents the \textsc{TraceSynth} pipeline, which consists of four stages:
(1) pre-processing of real data to transform raw kernel traces into structured, reproducible representations,
(2) probabilistic sequence modeling to capture recurring execution patterns and attribute correlations observed in real kernel traces,
(3) constraint-guided validation and repair to enforce kernel-level execution invariants, and
(4) downstream utility evaluation, where we assess the performance of synthetic traces when used to train diagnostic models.

These stages emphasize robustness, semantic correctness, and compatibility with downstream system-analysis workflows, while maintaining a predictable and reusable generation process.

\subsection{Data Preprocessing}
\label{sec:data-preprocessing}

This section describes the preprocessing pipeline that converts raw kernel traces into ML-ready representations.
\subsubsection{Trace Extraction and Formalization}
\label{sec:trace-extraction}

We process kernel execution traces captured using the LTTng framework~\cite{Desnoyers2006The}. Raw binary traces are decoded into a time-ordered sequence of kernel events
$X = \{e_1, e_2, \dots, e_T\}$, where each event $e_t$ corresponds to a kernel-level observation emitted by the tracing infrastructure. We assume only kernel-visible state and do not rely on external application-level context or hardware counters beyond what is recorded in the trace, to reflect common deployment constraints in production systems.

Each event $e_t$ records a high-resolution timestamp, an event type identifier, and associated metadata payloads. This decoding step mirrors common engineering practice when exporting kernel traces for offline analysis and debugging. A representative excerpt from an LTTng trace log is shown below:

\begin{verbatim}
[08:46:31.491346508] (+0.000000000) flutin kmem_kmalloc:
{ cpu_id = 1 }, { call_site = 0xFFFFFFFFA09A81A7,
  ptr = 0xFFFF880789E9B800, bytes_req = 739,
  bytes_alloc = 1024 }

[08:46:31.491349294] (+0.000002786) flutin kmem_kfree:
{ cpu_id = 1 }, { call_site = 0xFFFFFFFFA09A827A,
  ptr = 0xFFFF880789E9B800 }

[08:46:31.491349830] (+0.000000536) flutin sched_waking:
{ cpu_id = 1 }, { comm = "lttng-consumerd",
  tid = 2208, prio = 20, target_cpu = 2 }
\end{verbatim}

\subsubsection{Multi-Channel Event Representation}
\label{sec:multi-channel}

Rather than modeling only event types, we extract a multi-channel representation that captures both temporal and system-level context commonly available in kernel traces. Each kernel event is represented using six channels: 
(i) \texttt{event}---the event type identifier, 
(ii) \texttt{dt}---the inter-event time delta in seconds, 
(iii) \texttt{cpu}---the CPU core identifier, 
(iv) \texttt{tid}---the thread identifier, 
(v) \texttt{comm}---the process command name, and 
(vi) \texttt{ret}---the return value for system calls.

These channels capture complementary aspects of execution behavior, including event semantics (\texttt{event}, \texttt{ret}), timing dynamics (\texttt{dt}), scheduling context (\texttt{cpu}, \texttt{tid}), and workload identity (\texttt{comm}). Together, they allow the model to learn correlations observed in real kernel traces across time, execution context, and control flow.

The selected channels reflect metadata that is stable across executions and widely available in kernel-level tracing frameworks. In later sections, we explicitly evaluate how much each channel contributes to downstream utility through feature ablation studies.

\subsubsection{Vocabulary Construction and Structured Storage}
\label{sec:parquet}

To enable consistent numeric encoding across traces, we construct global vocabularies by scanning the training corpus and assigning deterministic integer identifiers to categorical values. All vocabularies include special tokens for padding and unknown values and are frozen prior to model training to ensure reproducibility across training, sampling, and evaluation.

The decoded traces are converted into Apache Parquet format~\cite{vohra2016apache} (a standard columnar storage format) for efficient columnar storage and access. Each trace run yields one Parquet file containing event-level records. This representation supports scalable preprocessing and enables efficient column-wise operations during windowing and feature extraction\cite{ivanov2020impact}.

While our method uses Parquet for efficiency, this storage step is independent of the underlying tracing framework. Any execution trace that can be decoded into a structured event table with timestamps and categorical attributes can be processed by the same pipeline, including traces produced by alternative kernel or system-level tracing infrastructures.

\subsubsection{Windowed Sequence Construction}
\label{sec:windowing}

Execution traces are long, variable-length event streams, whereas sequence models require fixed-length inputs. We therefore segment each trace into overlapping fixed-length windows using a sliding-window approach. Overlapping windows preserve temporal continuity across segments and allow recurring or rare execution patterns to appear in multiple training samples, improving robustness during training.

Event identifiers and CPU identifiers are preserved as categorical values. Inter-event time deltas (\texttt{dt}) are log-normalized using $\log(1 + \Delta t)$ to reduce scale variance and stabilize training. Thread identifiers are hashed into 256 buckets to control cardinality and memory footprint while preserving coarse scheduling structure. Categorical fields (\texttt{comm}, \texttt{ret}) are mapped using the frozen vocabularies defined earlier.

We generate windowed datasets at three sequence lengths ($L \in \{256, 1024, 4096\}$), corresponding to the context-length configurations evaluated in Section~\ref{sec:evaluation}. Each window contains aligned tensors for all six channels with shape $(N, L)$, where $N$ is the number of windows and $L$ is the sequence length. The resulting datasets are stored as compressed NumPy~\cite{harris2020array} archives (\texttt{.npz}) and serve as the machine-learning-ready input for all experiments. This windowing strategy applies to any temporally ordered execution trace and does not rely on kernel-specific semantics.

\subsection{Diffusion Model Architecture}
\label{sec:diffusion-model}

\textsc{TraceSynth} uses a Transformer-based denoising diffusion model to generate multi-channel execution traces. Although our evaluation focuses on kernel traces, the architecture operates over generic event sequences composed of discrete and continuous attributes. This design allows the same model to be applied to other system or application-level traces with similar structural properties.

Each windowed trace $\mathbf{x}$ of length $L$ contains six channels: \texttt{event}, \texttt{cpu}, \texttt{tid}, \texttt{comm}, \texttt{ret}, and \texttt{dt}. Discrete channels are embedded via learned tables $E_c \in \mathbb{R}^{|V_c| \times d}$, while the continuous time-delta channel is projected via an MLP~\cite{taud2017multilayer}. At each position, embeddings are summed to form a unified latent representation:

\[
\mathbf{h}_0[i] = \sum_{c \in \mathcal{C}_{\text{disc}}} E_c(x_c[i]) \;+\; \mathrm{MLP}_{dt}(x_{dt}[i]),
\quad \mathbf{h}_0 \in \mathbb{R}^{L \times d}.
\]

The forward diffusion process progressively corrupts the initial representation $\mathbf{h}_0$ with Gaussian noise over $T$ time steps~\cite{10.5555/3045118.3045358}:

\[
\mathbf{h}_t = \sqrt{\bar{\alpha}_t}\,\mathbf{h}_0 + \sqrt{1-\bar{\alpha}_t}\,\boldsymbol{\epsilon},
\quad \boldsymbol{\epsilon} \sim \mathcal{N}(0, I),
\]
where $\bar{\alpha}_t = \prod_{s=0}^{t}\alpha_s$ and $\{\alpha_t\}$ follow a linear $\beta$-schedule. This formulation enables the model to learn a smooth denoising trajectory over long event sequences while remaining agnostic to the specific semantics of individual event types.

The reverse process uses a Transformer encoder $\epsilon_{\theta}$ to predict the injected noise. Timestep conditioning is added via sinusoidal embeddings:

\[
\mathbf{t}_{\text{emb}} = \mathrm{MLP}_{\text{time}}(\mathrm{SinEmb}(t)), \quad
\hat{\boldsymbol{\epsilon}} = \epsilon_{\theta}(\mathbf{h}_t + \mathbf{t}_{\text{emb}}).
\]

The training objective combines a standard diffusion loss with an auxiliary reconstruction loss:
\[
\mathcal{L}_{\text{latent}} = \left\lVert \hat{\boldsymbol{\epsilon}} - \boldsymbol{\epsilon} \right\rVert_2^2.
\]
To maintain semantic consistency across channels, we decode the estimated clean state
$\hat{\mathbf{h}}_0 = (\mathbf{h}_t - \sqrt{1-\bar{\alpha}_t}\,\hat{\boldsymbol{\epsilon}}) / \sqrt{\bar{\alpha}_t}$ and optimize per-channel reconstruction:
\[
\mathcal{L}_{\text{recon}} = \sum_{c \in \mathcal{C}_{\text{disc}}} \mathrm{CE}(\hat{y}_c, y_c) + \mathrm{MSE}(\widehat{dt}, dt).
\]
The final objective is $\mathcal{L} = \mathcal{L}_{\text{latent}} + \lambda\,\mathcal{L}_{\text{recon}}$, where $\lambda$ weights the auxiliary term.

To generate synthetic traces, we initialize a latent sequence with Gaussian noise and iteratively apply the reverse diffusion process from $t=T-1$ to $0$. Discrete channels are decoded via argmax over logits, while continuous values are generated directly. The resulting synthetic sequences preserve the same multi-channel structure as the original traces and can be consumed by downstream analysis pipelines without modification.

\paragraph{Scope.}
The diffusion model operates on generic multi-channel event sequences and does not hard-code kernel-specific execution rules. Kernel traces serve as a concrete instantiation in this work, but the modeling formulation itself assumes only temporally ordered events with mixed discrete and continuous attributes.

\subsection{Training Procedure}
\label{sec:training}

Models are trained on windowed NPZ trace shards using the architecture, diffusion schedule, and loss formulation described above. The training procedure operates on generic windowed multi-channel event sequences and does not rely on kernel-specific supervision beyond the observed data distribution. For each mini-batch, a diffusion timestep is sampled uniformly, noise is injected according to the forward process, and the Transformer denoiser predicts the injected noise.

We train separate models at different sequence lengths ($L \in \{256, 1024, 4096\}$) to assess how temporal context affects generation quality. Batch sizes are adjusted to fit GPU memory constraints, while all other architectural and optimization hyper-parameters remain fixed across configurations to enable controlled comparison.

Optimization uses AdamW with cosine learning-rate decay. Model selection is based on validation loss computed on held-out windows drawn from the same distribution as the training data. The best-performing checkpoint for each configuration is retained and used consistently for sampling and downstream evaluation.

\subsection{Constraint Learning and Post-hoc Repair}
\label{sec:constraints-repair}

Neural sequence generators can produce locally plausible event sequences that violate global execution invariants observed in real traces. These violations arise because generative models optimize likelihood rather than semantic validity. While our evaluation focuses on kernel traces, the constraint-learning and repair mechanism applies more broadly to structured event sequences in which valid executions obey recurring temporal, structural, or attribute-level constraints.

Common violation classes include: (i) \emph{invalid event transitions} (e.g., system call exit without entry), (ii) \emph{temporal violations} (inter-event delays outside feasible ranges), and (iii) \emph{attribute inconsistencies} (events occurring on CPU cores or process contexts never observed for those event types). \textsc{TraceSynth} addresses these issues through constraint-based validation and repair using invariants mined directly from real traces rather than manually specified rules.

We extract four classes of constraints from real NPZ shards:  
(i) \textbf{Event transitions}—a directed graph $\mathcal{G} = (V, E)$ where $(e_i, e_j) \in E$ if $e_j$ follows $e_i$ in real traces;  
(ii) \textbf{Temporal bounds}—empirical minimum and maximum inter-event deltas per event type;  
(iii) \textbf{CPU affinity}—allowed CPU sets per event type;  
(iv) \textbf{Attribute validity}—allowed values for \texttt{tid}, \texttt{comm}, and \texttt{ret} conditioned on event type.  
All constraints are learned once from the full set of real traces across workloads and stored as a universal reference, enabling consistent validation across models and sampling configurations.

\paragraph{Constraint-based distance metrics.}
We quantify synthetic trace validity using four distance metrics. \emph{Transition distance} measures invalid event pairs:
\[
D_{\text{trans}}(\hat{X}) = 1 - \frac{1}{|\hat{X}|-1} \sum_{t} \mathbb{I}\left[(\hat{e}_t, \hat{e}_{t+1}) \in \mathcal{G}\right].
\]
\emph{Temporal distance} measures timing violations:
\[
D_{\text{time}}(\hat{X}) = \frac{1}{|\hat{X}|} \sum_{t} \mathbb{I}\left[\Delta t_t \notin [\min_e, \max_e]\right].
\]
\emph{CPU affinity distance} measures invalid CPU assignments:
\[
D_{\text{cpu}}(\hat{X}) = \frac{1}{|\hat{X}|} \sum_{t} \mathbb{I}\left[\hat{cpu}_t \notin \mathcal{C}_{\hat{e}_t}\right].
\]
\emph{Attribute validity distance} aggregates categorical violations:
\[
D_{\text{attr}}(\hat{X}) = \frac{1}{|\hat{X}|} \sum_{t} \mathbb{I}\left[\exists a \in \mathcal{A} : \hat{a}_t \notin \mathcal{V}_{\hat{e}_t}^{(a)}\right].
\]

\paragraph{Post-hoc repair.}
For each detected violation, the offending attribute is replaced by sampling from the corresponding allowed set conditioned on event type. Repair is applied uniformly as a post-processing step, which decouples generative modeling from semantic validation and allows different generators to reuse the same constraint reference. Distances are reported both before and after repair to verify invariant satisfaction and quantify the effectiveness of the repair mechanism.

\section{Evaluation Setup}
\label{sec:evaluation}

This section evaluates the effectiveness, robustness, and cost--quality trade-offs of \textsc{TraceSynth}. Our evaluation is structured around four research questions designed to assess when synthetic execution traces can safely augment or substitute limited real trace data in practical system-analysis pipelines.

\begin{itemize}[leftmargin=*]
  \item \textbf{RQ1:} When can synthetic execution traces safely augment limited real training data?
  \item \textbf{RQ2:} Does constraint-guided repair consistently improve synthetic data quality?
  \item \textbf{RQ3:} How does increasing diffusion model context length affect synthetic data quality?
  \item \textbf{RQ4:} Can diffusion models with reduced feature complexity achieve comparable quality with lower computational cost?
\end{itemize}

All research questions are evaluated using a controlled downstream next-event prediction task on held-out real execution traces, where training conditions and model configurations are held constant to isolate the effect of synthetic data. While next-event prediction is not an end application by itself, it serves as a rigorous proxy for evaluating whether models trained on synthetic data capture realistic temporal dependencies, rare-event structure, and execution patterns relied upon by downstream diagnostic and trace-driven tools.

To facilitate reproducibility and independent validation, all implementations and experimental configurations are publicly available.\footnote{\url{https://github.com/17YuvrajSehgal/SyntheticLogGeneration}}

\subsection{Compute Environment and Datasets}

All experiments were executed on the Nibi cluster, a system administered by SHARCNET and the Digital Research Alliance of Canada.\footnote{\url{https://alliancecan.ca}} Each training and sampling job was allocated a single NVIDIA H100 GPU (80\,GB), 8 CPU cores, and 32\,GB of host memory. Mixed-precision training with \texttt{bf16} and TF32 acceleration was enabled to maximize throughput and ensure stable optimization. While we use H100-class hardware for experimental efficiency and scalability, the proposed pipeline and modeling approach are not tied to a specific GPU architecture.

\paragraph{Datasets and Industrial Relevance.}
We evaluate \textsc{TraceSynth} using the LTTng kernel trace dataset introduced by Martin et al.~\cite{martin2017lttng}. We focus on six benchmarks from the Phoronix Test Suite (\emph{ffmpeg}, \emph{iozone}, \emph{pybench}, \emph{scimark2}, \emph{stream}, \emph{unpack-linux}), spanning compute-intensive, memory-intensive, and I/O-heavy workloads. Each benchmark provides 32 independent execution runs, ensuring sufficient diversity for training and evaluation.

These workloads were selected in consultation with Ciena engineers as representative proxies for the scheduling behavior, resource contention, and execution patterns commonly observed in proprietary network control-plane software. By relying on standard, open benchmarks, we ensure full reproducibility while capturing realistic kernel-level dynamics relevant to industrial systems, without exposing sensitive production traces or violating data-retention policies.

\subsection{Model and Training Configuration}
\label{sec:model-config}

All experiments use the same diffusion-model architecture to isolate the effects of temporal context length and feature availability. The denoising network is a Transformer encoder with embedding dimension $d_{\text{model}}=256$, 4--8 attention heads, and 4--8 layers depending on configuration; these variations are applied only where explicitly evaluated (RQ3 and RQ4) and are not jointly tuned. The diffusion process uses $T=1000$ time steps with a linear noise schedule from $\beta_{\text{start}}=10^{-4}$ to $\beta_{\text{end}}=0.02$.

Models are optimized using AdamW~\cite{loshchilov2017decoupled} with an initial learning rate of $2 \times 10^{-4}$ and cosine decay. Aside from the experimental variables described below, all architectural and optimization hyper-parameters are held constant across runs, and no configuration is tuned on a per-RQ basis. Table~\ref{tab:hyperparams} summarizes the shared configuration.

\begin{table}[t]
\centering
\caption{Model and training hyper-parameters (shared across all experiments).}
\label{tab:hyperparams}
\begin{tabular}{ll}
\toprule
\textbf{Parameter} & \textbf{Value} \\
\midrule
Model type & Transformer-based diffusion model \\
Embedding dimension ($d_{\text{model}}$) & 256 \\
Number of Transformer layers & 4--8 \\
Number of attention heads & 4--8 \\
Diffusion steps ($T$) & 1000 \\
Noise schedule & Linear $\beta$-schedule \\
Optimizer & AdamW \\
Learning rate & $2 \times 10^{-4}$ \\
Learning-rate schedule & Cosine decay \\
Precision & bf16 (TF32 enabled) \\
\bottomrule
\end{tabular}
\end{table}

\subsection{Sampling and Generation}

For each trained diffusion model, we generate synthetic kernel traces using DDIM sampling with 50 denoising steps starting from Gaussian noise. This configuration is held constant across all experiments and reflects a practical trade-off between sample quality and generation cost. For each model configuration, we generate 10{,}000 synthetic samples and store them in the same windowed NPZ format as real traces, ensuring identical downstream preprocessing.

\subsection{Downstream Task Evaluation}
\label{sec:downstream-eval}

We evaluate synthetic trace utility using a downstream next-event prediction task, which serves as a controlled proxy for assessing whether models trained on synthetic data can generalize to real kernel execution behavior. We treat next-event prediction as a conservative proxy task; failure to preserve temporal and rare-event structure in this setting typically compounds in higher-level tasks such as anomaly detection or root cause analysis.

\paragraph{Task Definition.}
Given a sequence of $L_{\text{pred}}=128$ consecutive kernel events, the model predicts the next event type (384-way classification). This task requires modeling both local event patterns and longer-range system dependencies.

\paragraph{Predictor Architecture.}
The downstream predictor is a Transformer-based classifier with $d_{\text{model}}=256$, 8 attention heads, and 4 layers. For multi-channel inputs, features are embedded separately, concatenated, and fused via a linear projection. The final hidden state is passed through a two-layer MLP to produce event logits.

\paragraph{Training Protocol.}
Predictors are trained for up to 20 epochs using AdamW ($\text{lr}=10^{-4}$) with cross-entropy loss. Training employs sliding windows with stride 64, batch size 64, and early stopping with patience 3 based on validation macro-F1.

\paragraph{Training Set Configurations (RQ1 \& RQ2).}
To evaluate augmentation effectiveness (RQ1) and repair utility (RQ2), predictors are trained under three conditions:
(i) \textbf{Real-only} (baseline),
(ii) \textbf{Combined (Repaired)}: a 50/50 mix of real and constraint-repaired synthetic traces, and
(iii) \textbf{Combined (No Repair)}: a 50/50 mix of real and unrepaired synthetic traces.
All predictors are evaluated on the same held-out real test set.

\paragraph{Ablation Study Configurations (RQ3 \& RQ4).}
To assess scalability (RQ3) and complexity trade-offs (RQ4), we train additional diffusion models varying two dimensions:
(i) \textbf{Context Length}: sequences of length $L \in \{256, 1024, 4096\}$, and
(ii) \textbf{Feature Richness}: three channel subsets ranging from \textbf{Base} (Event+Time only) to \textbf{Full} (all 6 channels).
For these studies, downstream predictors are instantiated with matching context lengths and input channels to isolate the impact of the generative model's capacity.

\subsection{Evaluation Metrics}
\label{sec:evaluation-metrics}

We evaluate downstream predictor performance using complementary metrics that capture both rare-event behavior and overall prediction quality. All metrics are computed on the same held-out real test set to ensure consistent comparison across training configurations.

\paragraph{F1-Score (Macro).}
Our primary metric is macro-averaged F1-score, which assigns equal weight to all event classes. Kernel traces are highly imbalanced, with frequent scheduler events dominating execution while rare events (e.g., error paths or I/O corner cases) are often operationally critical. Macro-F1 prevents models from achieving inflated scores by over predicting frequent events and therefore serves as our main comparison metric.

\paragraph{F1-Score (Weighted).}
Weighted F1-score accounts for class frequency and reflects performance on common execution paths. We report it as a complementary measure to macro-F1.

\paragraph{Accuracy.}
Classification accuracy is reported for completeness but is not interpreted in isolation due to severe class imbalance.

\paragraph{Top-$K$ Accuracy.}
We report Top-5 and Top-10 accuracy to capture plausibility under execution nondeterminism, where multiple next events may be valid due to scheduling or I/O interleaving.

\section{Results}
\label{sec:results}

We organize results around four research questions (RQ1--RQ4) that assess when and how synthetic kernel traces can be safely used in place of real data. All results are based on a downstream next-event prediction task evaluated on a held-out real test set. We report macro-F1 as the primary metric to capture rare-event behavior, with accuracy and Top-K accuracy as secondary indicators of overall plausibility.

\subsection{RQ1: When can synthetic kernel traces safely augment limited real training data?}
\label{sec:rq1}

To evaluate the utility of synthetic data augmentation, we simulate a common industrial scenario: data scarcity. Collecting large-scale production traces is often constrained by operational overhead or privacy policies. We therefore assess whether synthetic traces can effectively double the dataset size without degrading downstream model quality.

We compare (i) \textbf{Real-only} training against (ii) \textbf{Combined} training (50\% real + 50\% constraint-repaired synthetic). Table~\ref{tab:rq1} presents the results, where bold values indicate negligible degradation ($<5\%$) and italics denote moderate degradation ($5\text{--}20\%$).

Results in Table~\ref{tab:rq1} reveal strong workload dependence. Compute-heavy and structurally regular workloads benefit most from augmentation, while I/O-intensive workloads exhibit larger degradation. At $L=4096$, \texttt{scimark2} achieves 87.2\% macro-F1 with synthetic augmentation, only 2.6 percentage points below the real-only baseline (89.8\%). In contrast, I/O-intensive workloads such as \texttt{stream} and \texttt{iozone} exhibit substantial degradation (25--36\%), indicating that synthetic traces are less reliable for workloads dominated by asynchronous I/O and complex kernel interactions under the evaluated configurations.

Increasing diffusion context length consistently reduces degradation. Average macro-F1 loss decreases from 38.1\% at $L=256$ to 17.7\% at $L=4096$, suggesting that long-range temporal context is critical for preserving rare-event behavior in synthetic traces.

Secondary metrics in Table~\ref{tab:rq1-secondary} provide additional insight. While macro-F1 exposes degradation on rare events, weighted F1 and accuracy remain high (86--97\%), and Top-5/Top-10 accuracy exceeds 96\% across all configurations. This indicates that models trained with synthetic data perform well on frequent events and often include the correct next event among a small set of plausible alternatives, even when the top prediction is incorrect.

\noindent\fbox{%
    \parbox{0.95\linewidth}{%
        \textbf{Industrial Takeaway:} Synthetic augmentation is production-ready for compute-bound or structured workloads (e.g., numerical processing), provided sufficient context length ($L \ge 4096$). For complex, asynchronous I/O workloads (e.g., high-throughput networking), synthetic data should be used with caution, as it may not yet fully capture the long-tail variance of real production traffic.
    }%
}

\begin{table*}[t]
\centering
\caption{RQ1: Performance trade-offs when doubling the training dataset size using synthetic data. We compare training on real data (Real-only) with training on data composed of 50\% real and 50\% synthetic traces (Combined). $\Delta$F1 reports the change in macro-F1 score introduced by synthetic augmentation across workloads and context lengths.}
\label{tab:rq1}
\setlength{\tabcolsep}{4pt}
\begin{tabular}{@{}lccccccccc@{}}
\toprule
\multirow{2}{*}{\textbf{Benchmark}} & \multicolumn{3}{c}{\textbf{L=256}} & \multicolumn{3}{c}{\textbf{L=1024}} & \multicolumn{3}{c}{\textbf{L=4096}} \\
\cmidrule(lr){2-4} \cmidrule(lr){5-7} \cmidrule(lr){8-10}
& Real & Combined & $\Delta$F1 & Real & Combined & $\Delta$F1 & Real & Combined & $\Delta$F1 \\
\midrule
ffmpeg       & 69.9\% & 32.0\% & -37.9\% & 82.9\% & 60.1\% & -22.8\% & 81.5\% & 64.4\% & -17.1\% \\
iozone       & 64.0\% & 19.9\% & -44.1\% & 67.7\% & 34.8\% & -32.9\% & 69.3\% & 40.8\% & -28.5\% \\
pybench      & 70.6\% & 41.8\% & -28.8\% & 89.6\% & 69.7\% & -19.9\% & 88.6\% & \textit{78.3\%} & \textit{-10.3\%} \\
scimark2     & 72.0\% & 40.6\% & -31.4\% & 88.5\% & 68.0\% & -20.5\% & 89.8\% & \textbf{87.2\%} & \textbf{-2.6\%} \\
stream       & 68.5\% & 17.6\% & -50.9\% & 70.5\% & 40.7\% & -29.8\% & 69.7\% & 44.9\% & -24.8\% \\
unpack-linux & 63.4\% & 27.8\% & -35.6\% & 69.1\% & 44.3\% & -24.8\% & --- & 43.8\% & --- \\
\midrule
\textbf{Average} & \textbf{68.1\%} & \textbf{30.0\%} & \textbf{-38.1\%} & \textbf{78.0\%} & \textbf{52.9\%} & \textbf{-25.1\%} & \textbf{79.8\%} & \textbf{59.9\%} & \textbf{-17.7\%} \\
\bottomrule
\end{tabular}
\end{table*}

\begin{table*}[t]
\centering
\caption{RQ1 (Secondary Metrics): Weighted F1, accuracy, and Top-$K$ accuracy for the Combined (50\% real + 50\% synthetic) configuration across workloads and context lengths.}
\label{tab:rq1-secondary}
\setlength{\tabcolsep}{3.5pt}
\begin{tabular}{@{}lcccccccccccc@{}}
\toprule
\multirow{2}{*}{\textbf{Benchmark}} & \multicolumn{4}{c}{\textbf{L=256}} & \multicolumn{4}{c}{\textbf{L=1024}} & \multicolumn{4}{c}{\textbf{L=4096}} \\
\cmidrule(lr){2-5} \cmidrule(lr){6-9} \cmidrule(lr){10-13}
& F1-W & Acc & Top-5 & Top-10 & F1-W & Acc & Top-5 & Top-10 & F1-W & Acc & Top-5 & Top-10 \\
\midrule
ffmpeg       & 85.9\% & 86.6\% & 95.8\% & 97.4\% & 91.9\% & 92.1\% & 98.6\% & 99.2\% & 93.8\% & 93.9\% & 99.4\% & 99.7\% \\
iozone       & 84.4\% & 84.7\% & 95.2\% & 96.9\% & 89.6\% & 89.7\% & 98.2\% & 99.1\% & 92.8\% & 92.9\% & 99.3\% & 99.6\% \\
pybench      & 87.4\% & 87.8\% & 95.2\% & 96.6\% & 94.2\% & 94.3\% & 98.6\% & 99.2\% & 96.1\% & 96.2\% & 99.6\% & 99.8\% \\
scimark2     & 87.0\% & 87.5\% & 95.1\% & 96.5\% & 93.8\% & 93.8\% & 98.5\% & 99.1\% & 96.9\% & 97.0\% & 99.7\% & 99.8\% \\
stream       & 84.0\% & 84.5\% & 98.0\% & 98.5\% & 88.3\% & 88.4\% & 99.2\% & 99.5\% & 89.8\% & 89.9\% & 99.6\% & 99.8\% \\
unpack-linux & 85.3\% & 85.6\% & 95.1\% & 96.8\% & 90.5\% & 90.6\% & 98.2\% & 99.0\% & 92.9\% & 93.0\% & 99.3\% & 99.7\% \\
\midrule
\textbf{Average} & \textbf{85.7\%} & \textbf{86.1\%} & \textbf{95.7\%} & \textbf{97.1\%} & \textbf{91.4\%} & \textbf{91.5\%} & \textbf{98.5\%} & \textbf{99.2\%} & \textbf{93.7\%} & \textbf{93.8\%} & \textbf{99.5\%} & \textbf{99.7\%} \\
\bottomrule
\end{tabular}
\end{table*}

\subsection{RQ2: Does constraint-guided repair consistently improve synthetic data quality?}
\label{sec:rq2}

To evaluate the effectiveness of constraint-guided repair as a post-processing step, we compare downstream performance under two  conditions: 
(i) \textbf{Combined (No Repair)}, which mixes 50\% real data with 50\% raw synthetic traces, and 
(ii) \textbf{Combined (Repaired)}, which replaces raw synthetic traces with their constraint-repaired counterparts. 
This comparison isolates the impact of repair while holding the amount of synthetic augmentation constant.

Table~\ref{tab:rq2} summarizes the effect of constraint-guided repair across workloads and context lengths; bold values indicate cases where repair improves macro-F1 by more than 2\%, italics denote improvements between 0.5--2\%, and remaining values indicate changes below 0.5\%. Across 12 out of 15 benchmark–context combinations (80\%), repair increases macro-F1 by 0.3--4.3\%. The largest gains occur at shorter context lengths ($L=256$), with improvements of +4.3\% for \texttt{scimark2} and +4.1\% for \texttt{pybench}. These settings correspond to regimes where diffusion models are more likely to violate kernel execution invariants due to limited temporal context. 

As context length increases, the marginal benefit of repair diminishes. 
At $L=4096$, improvements are near-neutral once a single anomalous case (\texttt{unpack-linux}), which dominates the aggregate average, is excluded. Manual inspection suggests this behavior is trace-specific rather than systematic.
This trend suggests that diffusion models trained with long temporal context implicitly learn many structural invariants, reducing reliance on explicit repair.

Importantly, constraint-guided repair rarely degrades performance. Aside from minor fluctuations for \texttt{ffmpeg} and a single anomalous result for \texttt{unpack-linux} at $L=4096$, repair does not introduce systematic harm. Overall, these results indicate that constraint-guided repair functions as a low-risk safety mechanism that is most beneficial when generative models operate under limited context or higher uncertainty.

\noindent\fbox{%
    \parbox{0.95\linewidth}{%
        \textbf{Industrial Takeaway:} Constraint-guided repair acts as a safety net. It is essential when generating traces under high uncertainty (short context) but becomes optional as model capacity and context length increase.
    }%
}

\begin{table*}[t]
\centering
\caption{RQ2: Effect of constraint-guided repair across benchmarks and context lengths. We compare Combined (No Repair) and Combined (Repaired) configurations. $\Delta$F1 reports the change in macro-F1 score introduced by applying constraint-guided repair.}
\label{tab:rq2}
\setlength{\tabcolsep}{4pt}
\begin{tabular}{@{}lccccccccccccc@{}}
\toprule
\multirow{2}{*}{\textbf{Benchmark}} & \multicolumn{4}{c}{\textbf{L=256}} & \multicolumn{4}{c}{\textbf{L=1024}} & \multicolumn{4}{c}{\textbf{L=4096}} \\
\cmidrule(lr){2-5} \cmidrule(lr){6-9} \cmidrule(lr){10-13}
& No Rep. & Repaired & $\Delta$F1 & Rel. & No Rep. & Repaired & $\Delta$F1 & Rel. & No Rep. & Repaired & $\Delta$F1 & Rel. \\
\midrule
ffmpeg       & 33.2\% & 32.0\% & -1.2\% & -3.6\% & 60.2\% & 60.1\% & -0.1\% & -0.2\% & 65.6\% & 64.4\% & \textit{-1.2\%} & -1.8\% \\
iozone       & 19.5\% & 19.9\% & \textbf{+0.4\%} & \textbf{+2.0\%} & 35.0\% & 34.8\% & -0.2\% & -0.6\% & 41.3\% & 40.8\% & \textit{-0.5\%} & -1.2\% \\
pybench      & 40.1\% & 41.8\% & \textbf{+1.6\%} & \textbf{+4.1\%} & 69.7\% & 69.7\% & +0.0\% & +0.0\% & 78.0\% & 78.3\% & +0.3\% & +0.3\% \\
scimark2     & 38.9\% & 40.6\% & \textbf{+1.7\%} & \textbf{+4.3\%} & 67.7\% & 68.0\% & +0.3\% & +0.4\% & 87.0\% & 87.2\% & +0.2\% & +0.3\% \\
stream       & 17.2\% & 17.6\% & +0.3\% & \textit{+1.8\%} & 39.5\% & 40.7\% & \textbf{+1.2\%} & \textbf{+3.1\%} & 44.2\% & 44.9\% & \textit{+0.7\%} & \textit{+1.7\%} \\
unpack-linux & 27.4\% & 27.8\% & +0.4\% & \textit{+1.4\%} & 43.9\% & 44.3\% & +0.4\% & \textit{+1.0\%} & 58.0\% & 43.8\% & -14.2\%* & -24.6\%* \\
\midrule
\textbf{Average} & \textbf{29.4\%} & \textbf{30.0\%} & \textbf{+0.5\%} & \textbf{+1.5\%} & \textbf{52.7\%} & \textbf{52.9\%} & \textbf{+0.3\%} & \textbf{+0.6\%} & \textbf{62.4\%} & \textbf{59.9\%} & \textbf{-2.5\%} & \textbf{-4.2\%} \\
\bottomrule
\end{tabular}
\vspace{2pt}
{\raggedright *\,Anomaly in unpack-linux $L{=}4096$; isolated outlier likely due to dataset or trace-specific irregularities.\par}
\end{table*}

\subsection{RQ3: How does increasing diffusion model context length improve synthetic data quality?}
\label{sec:rq3}

To assess the impact of temporal context on synthetic data quality, we compare downstream task performance when training on \textbf{Combined (Repaired)} datasets generated by diffusion models with different context lengths, $L \in \{256, 1024, 4096\}$. This comparison isolates the effect of context length while holding constant the use of synthetic augmentation and constraint-guided repair.

Table~\ref{tab:rq3} reports macro-F1 scores across six benchmarks and three context lengths, along with the absolute and relative improvements from $L=256$ to $L=4096$.

Overall, increasing context length substantially improves synthetic data quality across all workloads. On average, macro-F1 increases by 29.9 percentage points, corresponding to a 104\% relative improvement when moving from $L=256$ to $L=4096$. 

Structured, compute-heavy workloads exhibit the largest improvements. For \texttt{scimark2}, macro-F1 increases from 40.6\% at $L=256$ to 87.2\% at $L=4096$ (+46.6\%, +115\% relative), approaching real-only performance (89.8\%). Similarly, \texttt{pybench} improves by +36.5\% (+87\% relative), and \texttt{ffmpeg} improves by +32.3\% (+101\% relative). These results indicate that longer temporal context enables diffusion models to better capture recurring execution patterns and rare-event structure in deterministic workloads.

I/O-intensive workloads also benefit from increased context length, though to a lesser extent. \texttt{stream} improves by +27.4\% (+156\% relative), \texttt{iozone} by +20.9\% (+105\%), and \texttt{unpack-linux} by +16.0\% (+57\%). While these workloads remain more challenging due to asynchronous behavior and complex kernel interactions, longer context substantially reduces the performance gap observed at shorter sequence lengths.

We observe diminishing returns as context length increases. The average improvement from $L=256$ to $L=1024$ is +22.9 percentage points, whereas the additional gain from $L=1024$ to $L=4096$ is +7.0 points. This suggests that $L=4096$ represents a practical upper bound under current hardware constraints, beyond which additional context yields smaller marginal benefits.

\noindent\fbox{%
    \parbox{0.95\linewidth}{%
        \textbf{Industrial Takeaway:} Maximize context length ($L=4096$) whenever hardware permits to capture long-range dependencies. If GPU memory is limited, $L=1024$ offers a cost-effective middle ground, but $L=256$ is insufficient for capturing kernel dynamics.
    }%
}

\begin{table}[t]
\centering
\caption{RQ3: Effect of diffusion model context length on synthetic data quality.
All results use the Combined (Repaired) configuration.
$\Delta$F1 denotes the absolute macro-F1 change from $L=256$ to $L=4096$, and Rel. Gain the corresponding relative improvement.}

\label{tab:rq3}
\setlength{\tabcolsep}{4pt}
\begin{tabular}{@{}lccccc@{}}
\toprule
\textbf{Benchmark} & \textbf{L=256} & \textbf{L=1024} & \textbf{L=4096} & \textbf{$\Delta$F1} & \textbf{Rel. Gain} \\
\midrule
ffmpeg       & 32.0\% & 60.1\% & 64.4\% & \textit{+32.3\%} & \textit{+101\%} \\
iozone       & 19.9\% & 34.8\% & 40.8\% & +20.9\% & +105\% \\
pybench      & 41.8\% & 69.7\% & 78.3\% & \textit{+36.5\%} & +87\% \\
scimark2     & 40.6\% & 68.0\% & 87.2\% & \textbf{+46.6\%} & \textbf{+115\%} \\
stream       & 17.6\% & 40.7\% & 44.9\% & +27.4\% & +156\% \\
unpack-linux & 27.8\% & 44.3\% & 43.8\% & +16.0\% & +57\% \\
\midrule
\textbf{Average} & \textbf{30.0\%} & \textbf{52.9\%} & \textbf{59.9\%} & \textbf{+29.9\%} & \textbf{+104\%} \\
\bottomrule
\end{tabular}
\end{table}

\subsection{RQ4: Can simpler diffusion models achieve comparable synthetic data quality with lower computational cost?}
\label{sec:rq4}

To assess the cost–quality trade-off of diffusion model complexity, we conduct a cross-model ablation study on three representative benchmarks (\texttt{ffmpeg}, \texttt{pybench}, and \texttt{scimark2}). We train diffusion models with three progressively richer channel configurations:
(i) \textbf{Base} (event + dt; 2 channels),
(ii) \textbf{System} (event + dt + cpu + tid; 4 channels), and
(iii) \textbf{Full} (event + dt + cpu + tid + comm + ret; 6 channels).

For each diffusion model, we train downstream predictors using all compatible channel subsets, yielding a cross-evaluation matrix that isolates the impact of diffusion-model feature richness from predictor feature choice.
We select \texttt{ffmpeg}, \texttt{pybench}, and \texttt{scimark2} as representative workloads spanning structured compute, mixed execution, and runtime-driven behavior; preliminary results showed that I/O-dominated workloads exhibit limited sensitivity to feature richness, making them less informative for feature ablation.

Table~\ref{tab:rq4} reports macro-F1 scores for all valid diffusion–predictor combinations. Across all benchmarks, the Base diffusion model achieves performance within 1--3\% of the Full model, despite using only event type and timing information. For \texttt{ffmpeg}, the Base model achieves the highest macro-F1 (61.8\%), matching or exceeding both the System (61.7\%) and Full (60.9\%) configurations. Similarly, for \texttt{pybench}, the Base model achieves the best overall performance (71.3\%), slightly outperforming the Full model (71.2\%).

Only \texttt{scimark2} exhibits a consistent benefit from richer diffusion inputs: the Full model reaches 69.4\% macro-F1 compared to 68.5\% for the Base configuration (+0.9\%). This workload is highly structured and compute-intensive, suggesting that additional semantic and scheduling features provide marginal gains when execution behavior is more deterministic.

Within each diffusion model, varying the downstream predictor’s channel configuration produces only 1--3\% variation in macro-F1. This indicates that diffusion-model feature selection has a larger impact on synthetic data quality than downstream predictor feature richness in this setting.

Overall, these results show that diffusion models operating only on event types and timing information can generate synthetic traces that are nearly as useful as those produced by feature-rich models, while requiring fewer input channels and lower computational overhead. Richer feature sets offer limited additional benefit and are primarily advantageous for highly structured workloads.

\noindent\fbox{%
    \parbox{0.95\linewidth}{%
        \textbf{Industrial Takeaway:} Do not over-engineer the generator. A lightweight 2-channel diffusion model reduces training cost and inference latency while delivering nearly all of the practical utility of a complex multi-channel model. Feature-rich models are only justified for highly deterministic, compute-bound systems.
    }%
}

\begin{table}[t]
\centering
\caption{RQ4: Cross-model ablation results (macro-F1 \%).
Rows correspond to diffusion model feature sets and columns to downstream predictor features. All results use Combined (Repaired) with $L=4096$.
\textbf{Bold} indicates the best configuration per benchmark; \textit{italic} indicates within 1\% of best.}

\label{tab:rq4}
\footnotesize
\setlength{\tabcolsep}{3pt}
\begin{tabular}{@{}llcccc@{}}
\toprule
\textbf{Benchmark} & \textbf{Diffusion Model} & \textbf{event} & \textbf{event+dt} & \textbf{event+dt+cpu+tid} & \textbf{all 6} \\
\midrule
\multirow{3}{*}{\textbf{ffmpeg}} 
& Base (2 ch) & 60.6\% & \textbf{61.8\%} & --- & --- \\
& System (4 ch) & 60.8\% & \textit{61.7\%} & 60.5\% & --- \\
& Full (6 ch) & 60.8\% & 60.9\% & 59.7\% & 58.9\% \\
\midrule
\multirow{3}{*}{\textbf{pybench}} 
& Base (2 ch) & \textbf{71.3\%} & 70.6\% & --- & --- \\
& System (4 ch) & 70.3\% & 70.9\% & \textit{71.0\%} & --- \\
& Full (6 ch) & 70.0\% & \textit{71.2\%} & \textit{71.2\%} & 70.6\% \\
\midrule
\multirow{3}{*}{\textbf{scimark2}} 
& Base (2 ch) & 67.9\% & 68.5\% & --- & --- \\
& System (4 ch) & 67.8\% & 65.5\% & 67.0\% & --- \\
& Full (6 ch) & 67.5\% & 68.9\% & \textit{68.8\%} & \textbf{69.4\%} \\
\bottomrule
\end{tabular}
\end{table}

\section{Discussion and Industrial Implications}
\label{sec:discussion}

\begin{figure}[t]
\centering
\includegraphics[width=0.95\columnwidth]{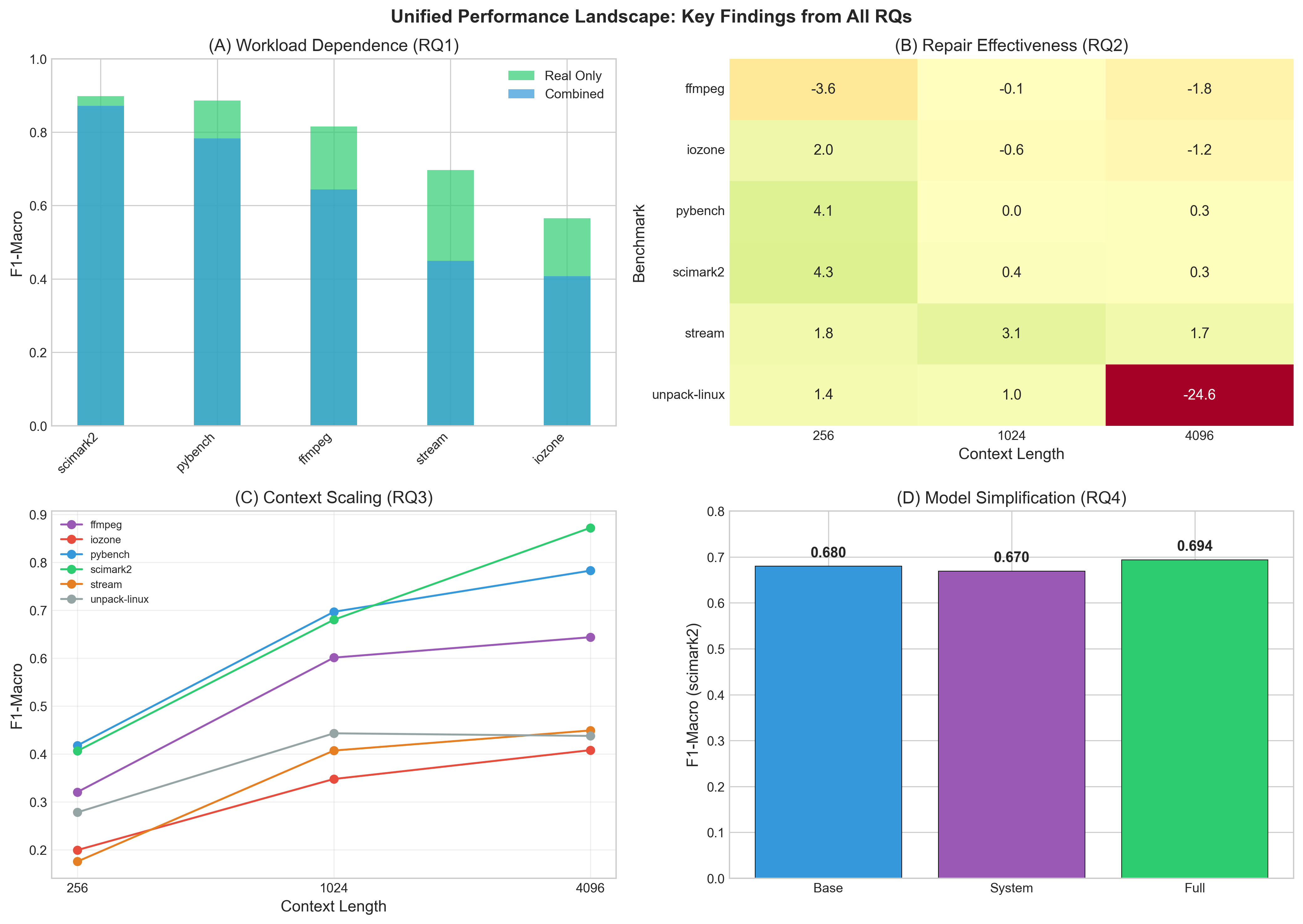}
\caption{Unified Performance Landscape. (A) Workload Dependence: Deterministic workloads (scimark2) achieve near-parity with real data. (B) Repair Effectiveness: Consistent gains at short context. (C) Context Scaling: The dominant quality driver ($L=4096$). (D) Model Simplification: Simple 2-channel models match complex ones.}
\label{fig:unified-landscape}
\end{figure}

We synthesize findings across our four research questions to provide practical guidance for deploying synthetic kernel traces in real-world ML pipelines. Figure~\ref{fig:unified-landscape} summarizes key trends across workloads, model configurations, and evaluation dimensions.

\subsection{Workload-Dependent Viability}
Our results show that the utility of synthetic kernel traces is fundamentally workload-dependent.

\begin{itemize}[leftmargin=*]
    \item \textbf{Deterministic workloads (ready):} Structured workloads like \texttt{scimark2} achieve near-parity with real data (87.2\% F1 vs. 89.8\% Real). For these systems, diffusion models successfully capture the main structure of execution.
    \item \textbf{I/O-heavy workloads (caution):} Workloads dominated by asynchronous interactions (\texttt{stream}, \texttt{iozone}) exhibit degradation even at long context lengths. The model struggles to infer external state (for example, disk queue depth) solely from the event log.
\end{itemize}
\textbf{Implication:} Practitioners can use synthetic augmentation with confidence for compute-bound control planes, and should treat it primarily as exploratory support when dealing with data-plane I/O traffic.

\subsection{The ROI of Context Length vs. Features}
Our ablation studies (RQ3 and RQ4) reveal a clear resource allocation strategy for industry: prioritize context over complexity. Increasing context from $L=256$ to $L=4096$ doubled model performance in our setting (approximately +104\% relative gain). In contrast, adding complex feature channels (CPU ID, Thread ID, command name, return code) yielded negligible gains (typically below 1\%).

\textbf{Implication:} Engineers should focus on maximizing temporal context and maintain minimal feature sets (event type and timing) where possible. Stripping inputs down to event and time allows the saved GPU memory to be reallocated to a larger context window.

\subsection{Practical Integration Scenarios}
Beyond simple augmentation, we identify three integration patterns where \textsc{TraceSynth} provides immediate value. TraceSynth is currently evaluated on open benchmarks selected in consultation with our industrial partner, Ciena, and is intended as a pre-deployment validation step before live production use.

\paragraph{1. Validating Trace Consumption Pipelines.}
Organizations maintain complex parsers and analytics pipelines. Synthetic traces are well suited for stress-testing these components. The high Top-$K$ accuracy (above 96\%) ensures that generated traces are structurally valid and plausible, which allows engineers to fuzz-test parsers and downstream components without collecting large volumes of production data.

\paragraph{2. Privacy-Preserving Development.}
Sharing production traces with external vendors or researchers is often blocked by compliance. Synthetic traces preserve the statistical behavior of the workload (for example, event frequency and temporal variability) without exposing sensitive payloads or customer identifiers. This enables safer cross-organization collaboration and experimentation when direct access to raw traces is restricted.

\paragraph{3. Augmenting Rare Behaviors.}
While the model struggles to predict rare events perfectly, it can replay them once they appear in the training data. By over-sampling windows that contain rare classes during generation, teams can create more balanced datasets for training classifiers that would otherwise be biased toward frequent scheduler events.

\subsection{Operational Guardrails}
Based on our experimental results, we provide the following operational guardrails for deploying generative models in production observability stacks:

\begin{tcolorbox}[breakable, colback=white, colframe=black, boxrule=0.4pt, arc=0pt, left=4pt, right=4pt, top=4pt, bottom=4pt]
\textbf{Do}
\begin{itemize}[leftmargin=*,nosep]
\item Use \textsc{TraceSynth} to augment data for deterministic, compute-heavy workloads.
\item Apply constraint-guided repair before downstream use.
\item Start with minimal feature sets (event + timing) to reduce training cost.
\item Use synthetic traces for fuzzing, testing, and lower-exposure data sharing.
\end{itemize}

\vspace{0.5em}

\textbf{Do not}
\begin{itemize}[leftmargin=*,nosep]
\item Use synthetic traces as the only data source for I/O-heavy anomaly detection.
\item Use synthetic data for safety-critical root cause analysis without validation.
\item Push context length past hardware limits; $L=4096$ was the best setting tested here.
\end{itemize}
\end{tcolorbox}

\subsection{Limitations and Threats to Validity}
\label{sec:limitations-and-threats-to-validity}

\textbf{Benchmark diversity.}
Our evaluation uses six Phoronix Test Suite benchmarks covering diverse workload types. While representative of common system behaviors, they do not capture user-driven dynamics or multi-tenant interference typical of large-scale production environments. Validation on proprietary or long-running production traces remains future work.

\textbf{Downstream task specificity.}
Synthetic data quality is evaluated via next-event prediction, a foundational task stressing temporal dependencies and rare-event structure. Other tasks (e.g., anomaly detection, performance forecasting) may exhibit different sensitivities to synthetic artifacts and are not evaluated here.

\textbf{Context length constraints.}
Experiments are limited to a maximum context length of $L=4096$ due to GPU memory constraints. Longer contexts may further improve generation quality, but cannot be assessed with current hardware.

\textbf{Model architecture scope.}
We focus on a Transformer-based diffusion model. Alternative architectures (e.g., state-space or convolutional) may offer different quality–efficiency trade-offs, so conclusions may not generalize to all generative designs.

\textbf{Repair strategy limitations.}
Constraint-guided repair uses greedy forward correction with learned transition rules. More advanced mechanisms could provide stronger guarantees at higher computational cost.

\noindent
Despite these limitations, the results for \texttt{scimark2} (87.2\% macro-F1, within 2.6\% of real data) demonstrate that diffusion-based synthetic trace generation is viable for structured workloads when context length and repair mechanisms are appropriately configured.

\section{Conclusion and Future Work}
\label{sec:conclusion}

High-fidelity kernel traces are a critical input for training AIOps and trace-driven system analysis models, yet their collection is often constrained by privacy requirements and production overhead. In this work, we presented \textsc{TraceSynth}, a diffusion-based framework for generating synthetic kernel execution traces. By combining long-context diffusion modeling with constraint-guided repair, we showed that synthetic traces can effectively augment limited real data for downstream machine learning tasks.

Our evaluation yields three key insights for industrial practice. First, context dominates: increasing temporal context to $L=4096$ is the strongest driver of quality and allows synthetic data to approach real-data performance for structured workloads. Second, simplicity scales: lightweight 2-channel models frequently match or slightly outperform feature-rich configurations, indicating that event ordering and timing capture most of the useful signal in kernel execution. Third, repair provides robustness: constraint-guided repair offers a low-cost mechanism for enforcing structural validity, making generative traces safer to integrate into analysis pipelines.

While \textsc{TraceSynth} performs well for structured compute workloads, highly asynchronous I/O-intensive systems remain challenging. Future work will focus on four directions:
(1) Industrial validation, including deployment on proprietary Ciena control-plane traces to assess behavior under live production conditions;
(2) State-aware modeling, by integrating explicit representations of system state to better capture I/O nondeterminism;
(3) Extended context, through linear-attention or state-space architectures that support longer sequences without quadratic memory costs; and
(4) Formal privacy guarantees, by incorporating differential privacy during training to support safe cross-organization trace sharing.

Together, these results demonstrate that diffusion-based synthetic trace generation is a practical tool for reducing data collection barriers in systems engineering, and support a shift from ad hoc tracing toward reusable, synthesized observability pipelines.

\balance
\bibliographystyle{ACM-Reference-Format}
\bibliography{references}

\end{document}